\documentclass{aastex62}

\graphicspath{{./}{}}

\received{February 25, 2019}
\revised{March  11, 2019}
\accepted{N/A}
\submitjournal{ApJ Letters}

\shorttitle{Parker Solar Probe Predictions}
\shortauthors{Riley et al.}

\begin{document}

\title{Predicting the Structure of the Solar Corona and Inner Heliosphere during Parker Solar Probe's First Perihelion Pass}

\correspondingauthor{Pete Riley}
\email{pete@predsci.com}

\author{Pete Riley, Cooper Downs, Jon A. Linker, Zoran Mikic, Roberto Lionello, and Ronald M. Caplan}
\affiliation{Predictive Science Inc. \\
9990 Mesa Rim Rd., Suite 170, \\
San Diego, CA 92121, USA}



\begin{abstract}

NASA's Parker Solar Probe (PSP) spacecraft reached its first perihelion of 35.7 solar radii on November 5th, 2018. To aid in mission planning, and in anticipation of the unprecedented measurements to be returned, in late October, we developed a three-dimensional magnetohydrodynamic (MHD) solution for the solar corona and inner heliosphere, driven by the then available observations of the Sun's photospheric magnetic field. Our model incorporates a wave-turbulence-driven (WTD) model to heat the corona. Here, we present our predictions for the structure of the solar corona and the likely {\it in situ} measurements that PSP will be returning over the next few months. We infer that, in the days prior to first encounter, PSP was immersed in wind emanating from a well-established, positive-polarity northern polar coronal hole. During the encounter, however, field lines from the spacecraft mapped to a negative-polarity equatorial coronal hole, within which it remained for the entire encounter, before becoming magnetically connected to a positive-polarity equatorial coronal hole. When the PSP data become available, these model results can be used to assist in their calibration and interpretation, and, additionally, provide a global context for interpreting the localized {\it in situ} measurements. In particular, we can identify what types of solar wind PSP encountered, what the underlying magnetic structure was, and how complexities in the orbital trajectory can be interpreted within a global, inertial frame. Ultimately, the measurements returned by PSP can be used to constrain current theories for heating the solar corona and accelerating the solar wind.

\end{abstract}



\section{Introduction} \label{sec:intro}

Parker Solar Probe (PSP) was launched on August 12, 2018. Through the use of seven gravity assists from Venus, it will, over a total of 24 orbits, reach a final heliocentric distance of 9.86 solar radii ($R_S$) \citep{fox16a}. At closest approach it will be traveling at speeds approaching 200 km/s, comparable to, or greater than the expected local speed of the solar wind at this distance. The primary scientific goals of the mission are to: (1) better understand what heats the solar corona and accelerates the solar wind; (2) determine the underlying structure and dynamics of the coronal magnetic field; and (3) better identify the mechanisms that accelerate and transport energetic particles in the corona \citep{fox16a}. 

PSP is carrying four instrument packages. FIELDS (Electromagnetic Fields Investigation)  consists of two flux-gate magnetometers, a search-coil magnetometer, and five plasma voltage sensors \citep{bale16a}. It measures electric and magnetic fields, as well as radio waves, Poynting flux, plasma density and electron  temperature. ISoIS (Integrated Science Investigation of the Sun) consists of two separate instruments: EPI-Hi and EPI-Lo, which measure energetic electrons, protons, and heavy ions \citep{mccomas16a}. SWEAP (Solar Wind Electrons Alphas and Protons) is composed of three instruments: two electrostatic analyzers and one Faraday cup, from which estimates of velocity, density, and temperature of electrons, protons, and alpha particles can be made \citep{kasper16a}. Finally, WISPR (Wide-field Imager for Solar Probe) comprises of two optical telescopes that provide coronagraph-like images of the corona and heliosphere \citep{vourlidas16a}. 

Global models of the solar corona and inner heliosphere are important tools for complementing any spacecraft mission \citep[e.g.][]{pizzo94a,linker99a,riley01a,aschwanden08a,torok18a}. However, this is particularly true for PSP, given its unique trajectory, narrow temporal windows for each encounter, and limited instrumentation. Most recently, we have developed a global, time-dependent MHD model, Magnetohydrodynamic Algorithm outside a Sphere (MAS), which includes the effects of waves and turbulence to  heat the corona, and the WKB approximation for wave pressures to accelerate the solar wind \citep{mikic18a}. Thus, it is particularly suitable for interpreting PSP measurements.  

In this report, we describe the results of MHD model solutions aimed at predicting what PSP will have observed around first encounter (31st October, 2018 to 11th November, 2018) during which time the spacecraft remains within 0.25 AU of the Sun. Although the measurements have already been made by the spacecraft, these simulations were completed prior to first encounter (October 30, 2018), and their interpretation is currently being made before the data have been publicly released. In this sense, it represents an informal prediction of the `future' state of the corona. In Section~\ref{sec:analytic}, we first provide an analytical summary of PSP's unique trajectory before reviewing the MHD solutions. We next present a global MHD model, and use the results to magnetically map PSP's trajectory back to the solar surface, predicting the likely origin of the solar wind plasma and magnetic field that PSP was immersed in during the first encounter. We then compare these model solutions with a Potential Field Source Surface (PFSS) model highlighting the differences between them. Finally, we fly the spacecraft though our heliospheric solutions producing a prediction of the plasma and magnetic field measurements that the FIELDS and SWEAP instruments will have made. We conclude by discussing the implications of this study, and anticipating how future encounters, complemented by measurements from Solar Orbiter, can provide additional constraints on the models, and lead to a better understanding of the physical processes at work. 

\section{Analytic Analysis} \label{sec:analytic}

\begin{figure}[ht!]
\plotone{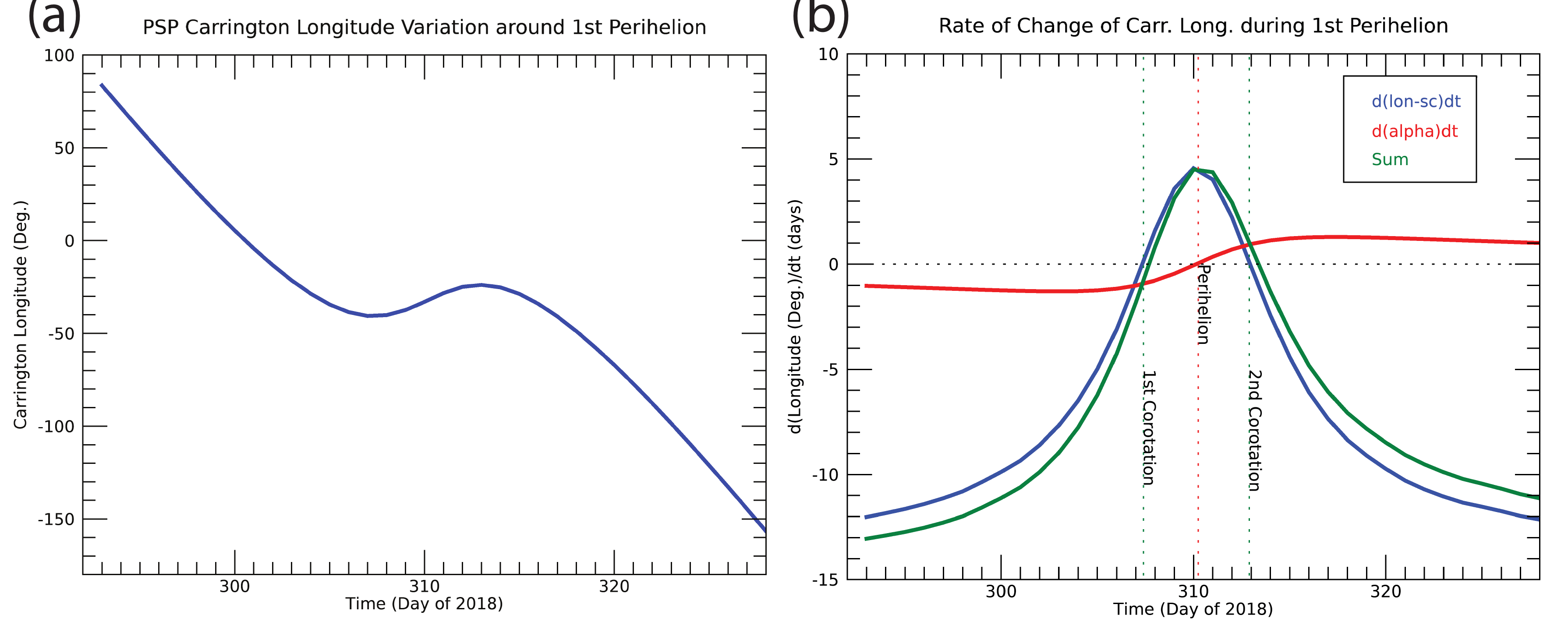}
\caption{(a) Variation of PSP's Carrington longitude position as a function of time during first encounter. (b) Variation of the time rate of change of PSP's Carrington longitude (blue), foot-print location based on its radial motion (red), and the sum of these two effects (green) as a function of time during first encounter.}
\label{fig-lon-dlon-v-time}
\end{figure}

Before discussing the MHD model, it is worth briefly considering some unique aspects of the PSP trajectory. Unlike many previous spacecraft, such as Wind, ACE, STEREO A and B, which traveled at an approximately constant rate in azimuth, because of its evolving eccentric orbit, PSP sweeps ever faster azimuthally as it approaches perihelion. This has implications for which magnetic field lines intercept the spacecraft, and, by extension, changes the inferred source region of the plasma being measured. Relative to Carrington coordinates, which is a fixed system rotating at a rate of 25.38 days,  the spacecraft moves backward in Carrington longitude until approximately the beginning of first encounter, when it corotates with the Sun. It then travels faster than the Sun's surface (at least that part that is traveling with an angular velocity of $2 \pi/25.38$ rads/day) briefly, before reaching a point of corotation again, and finally drifting further back in Carrington longitude (Figure~\ref{fig-lon-dlon-v-time} (a)). Between these two points of corotation lies perihelion,  at 35.7 $R_S$.

In addition to the direct effect of crossing solar meridians, an indirect effect occurs because of PSP's radial motion. As described by \citet{parker58a}, magnetic field lines convected out by the super-Alfv\'enic solar wind become increasingly wound up as the Sun rotates underneath them. This Archimedean pattern results in field lines of approximately $45^{\circ}$ at 1 AU for wind traveling at $\sim 450$ km s$^{-1}$. As the spacecraft accelerates into perilelion, it moves ever faster toward the Sun, which has the effect of causing it to sample these Archimedean field lines that connect progressively further to the east. This can be written as: 

\begin{equation}
\Delta \alpha = \frac{\Delta R (km)}{v (km/s)} \frac{360^{\circ}}{25.38 \times 86400 s}
\end{equation}
where $\Delta R$ is the change in heliocentric distance, and $v$ is the speed of the solar wind. This and the preceding effect are summarized in Figure~\ref{fig-lon-dlon-v-time} (b), which shows the time rate of change in the foot-point location of field lines connected to PSP due to its azimuthal motion (blue) and the radial motion (red). The former dominates, but, when combined (green), it has the overall effect of making the connectivity asymmetric, due to the fact that the radial motion is not symmetric around perihelion. Overall then, both first and second corotation are later in time with respect to perihelion, and, thus, the outward portion of the encounter lasts approximately one day longer than the inward portion, with respect to the points of corotation. 


\section{Modeling Approach} \label{sec:modeling}

In this study we used the MAS code, which solves the set of resistive MHD equations in spherical coordinates on a non-uniform mesh. The details of the model have been described elsewhere \citep[e.g.][]{mikic94a,riley01a,lionello01a,riley12a,caplan17b,mikic18a}. Here, we restrict our description to several relevant points. First, the model is driven by the observed photospheric magnetic field. We used HMI magnetograph observations from the SDO spacecraft to construct a boundary condition for the radial magnetic field at $1 R_S$ as a function of latitude and longitude. In particular, given the requirement of making the prediction prior to the start of first encounter, we built up a map based on observations during Carrington rotation (CR) 2208 and 2209. As a practical limitation, since PSP was effectively located at quadrature to Earth, during perihelion, this meant that the data lying under the spacecraft at the point of closest approach was approximately one week old (since it was observed from Earth, and rotated westward to build up the map). Additionally, due to inaccuracies stemming from projection effects, we applied a pole-fitting procedure to reconstruct the poorly observed polar regions, as described by \citet{mikic18a}. Second, the model is run in two stages: First the region from $1 - 30 R_S$ is modeled, followed by the region from $30 R_S$ to 1~AU, being driven directly by the results of the coronal calculation. Computationally, this approach is much more efficient, and, by overlapping the region between the simulations, we verified that the transition is seamless \citep{lionello13a}. Third, as noted above, this version of the model implements a Wave-Turbulence-Driven (WTD) approach for self-consistently heating the corona and invokes the WKB approximation for wave pressures, providing the necessary acceleration of the solar wind \citep{mikic18a}. The physical motivation for this heating model is that outward and reflecting Alfv\'en waves interact with one another, resulting in their dissipation, and heating of the corona \citep[e.g.][]{zank96a,verdini07a}. We have found that this approach can account for both the acceleration of solar wind along open field lines, as well as the heating of plasma entrained within closed-field regions \citep{lionello14a,downs16a}. 

\section{Results} \label{sec:results}

\begin{figure}[ht!]
\plotone{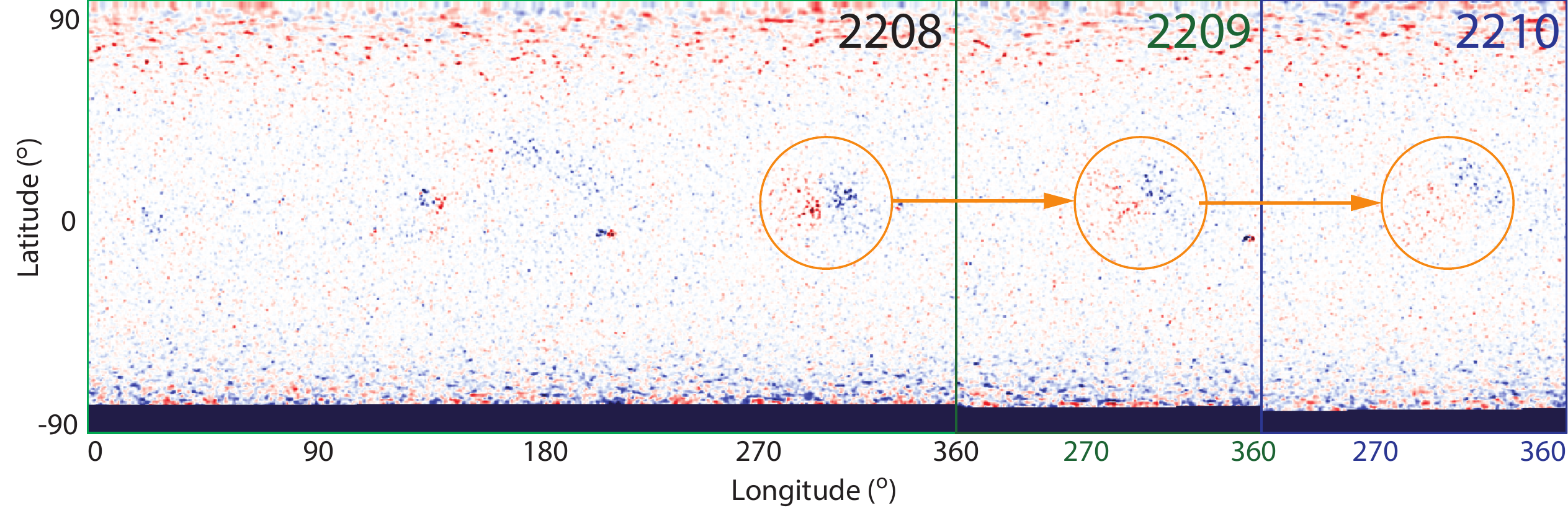}
\caption{Synoptic maps (Carrington longitude versus heliographic latitude) of the photospheric magnetic field for CR 2208, and parts of 2209 and 2210, showing the evolution of the active region at $\sim 330^{\circ}$ longitude.}
\label{fig-magnetograms}
\end{figure}

The boundary conditions used to drive the WTD coronal model are summarized in Figure~\ref{fig-magnetograms}. As described by \citet{mikic18a}, we multiplied the HMI-derived synoptic maps by a factor of 1.4 to account for the difference in magnetic field strengths measured by the HMI instrument as compared to   its predecessor, MDI onboard SOHO. As noted above, we initiated our production run one week prior to the beginning of first encounter, which required that we combine maps from CR 2208 and 2209. Since PSP was at quadrature relative to the Earth at Perihelion it it is not clear whether more timely data would have improved the accuracy of the solution. Retrospectively, however, it was useful to compare data near Carrington longitude $330^{\circ}$ for CRs 2208, 2209, and 2210 (Figure~\ref{fig-magnetograms}). During these rotations the active region lying under the spacecraft evolved significantly, becoming progressively weaker. This allows us, at least in a heuristic sense, to estimate likely uncertainties from a region that was neither well observed nor stationary in time. 

\begin{figure}[ht!]
\plotone{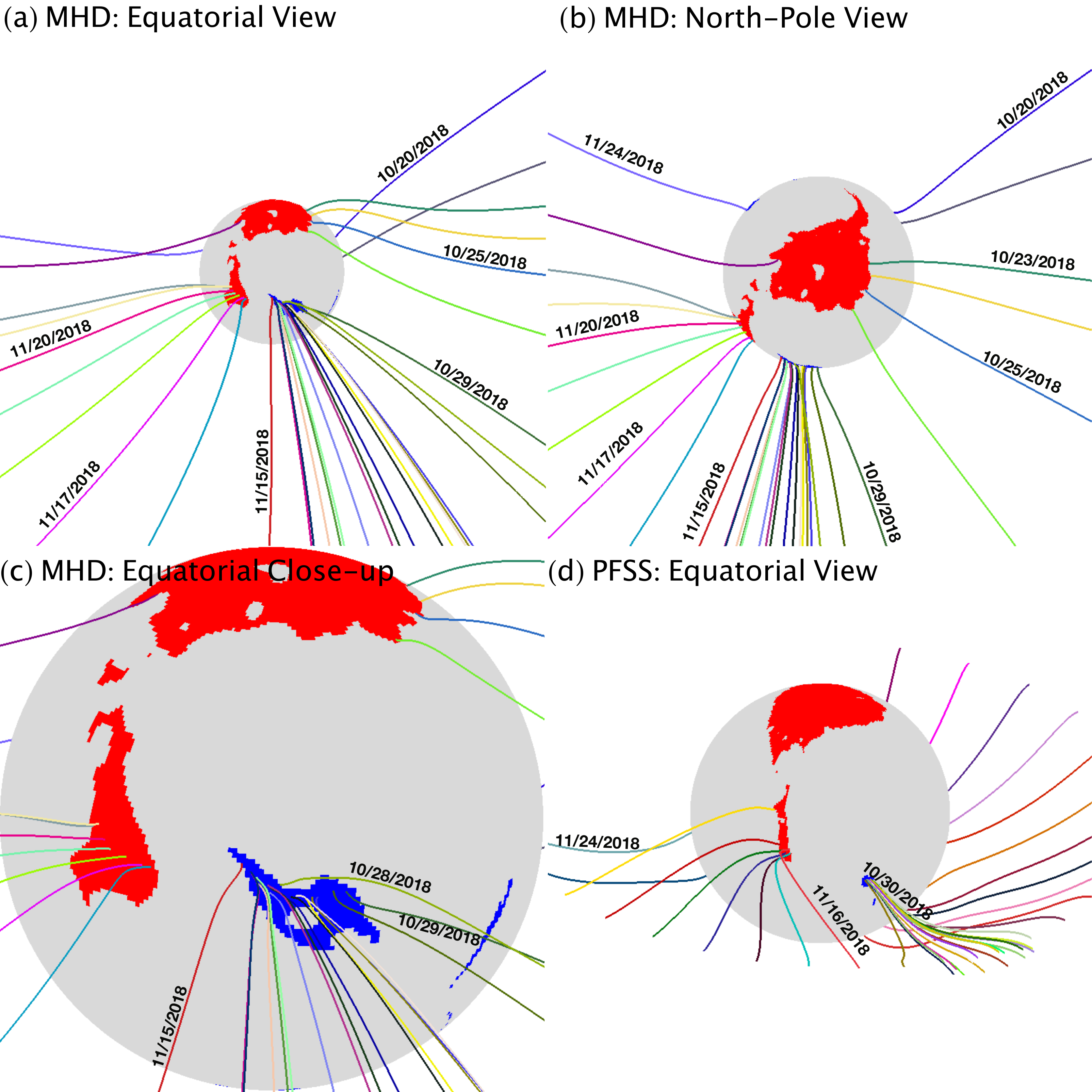}
\caption{Mapping of field lines from PSP's location in the inner heliosphere back to the solar surface. (a) Field lines (one per day) are shown mapped back to the coronal hole that they originated in using the MHD model solution. The underlying, observed polarity of the field is shown (red for radially-outward field lines and blue for inward field lines). (b) The same field lines are shown from a view above the north pole of the Sun. (c) An equatorial close-up view of the same results. (d) Similar field lines are mapped back to the surface using a PFSS model solution, where the spacecraft location has been first ballistically mapped back to 2.0 $R_S$.}
\label{fig-field-lines}
\end{figure}

The WTD-driven model was run using the aforementioned synoptic map out to 30 $R_S$. Then the output from this run was used to directly drive a second simulation from $29 R_S$ to 1 AU. Once complete, these data were interpolated onto a common grid and the orbital trajectory of PSP was flown though the results. For each day, over a week prior to, and following the encounter field lines at the spacecraft's location were traced back to the Sun. These are shown in Figure~\ref{fig-field-lines}(a-c). Each field line is drawn (in an arbitrary, but consistent colour) and a selection of them are stamped with the relevant date for that field line. We note that during the week prior to first encounter, the field lines all traced back to the positive-polarity northern polar coronal hole. Just prior to first encounter (10/29/2018), the connectivity to the spacecraft jumped to the negative-polarity equatorial coronal hole. PSP remained connected to this region for the entire first encounter, finally becoming connected with the positive-polarity, equatorial coronal hole on 11/16/2019. Several further points are worth noting. First, during first encounter, the field lines are bunched together (see panel (b), in particular). However, given the rapid acceleration of the spacecraft by this point, and the fact that it was reaching and exceeding corotation speeds, we cannot make any inferences on the expansion properties of the field lines. Second, during the central portion of the encounter, the field lines shown actually ``wrap around'' one another, indicating that the source region of plasma measured at PSP reversed toward the east, if only briefly. Again, this should be interpreted carefully: These are not field lines at a single point in time, but a sequence of field lines drawn on successive days. Third, initial connectivity with the equatorial hole is complex: On 10/28/2018, the field line connects to an eastward region of the CH, whereas one day later, it maps to the west-most portion of it. 

We also computed a PFSS solution using the same combined (CR 2208+2209) synoptic map. The resulting coronal hole structure and the mapping of the field lines intercepting PSP are shown in Figure~\ref{fig-field-lines} (d), and can be compared with the MHD results in panel (a). There are several similarities between the two solutions, although also some important distinctions. First, both solutions produce two equatorial coronal holes at roughly the same Carrington longitudes. Additionally, both solutions map to the same positive-polarity equatorial coronal hole for roughly the same intervals, spanning the first encounter.  However, the size and shape of the equatorial holes are substantially different, with the PFSS solution producing much smaller areas, in spite of the fact that we set the source surface radius to $2.0 R_S$, which is typically lower than is usually used, and chosen to open coronal holes as much as possible. Moreover, prior to first encounter, the PFSS solution mapped to the negative-polarity southern polar coronal hole, and, thus, this transition would not be associated with a current sheet crossing.  

\begin{figure}[ht!]
\centering
\includegraphics[scale=0.25]{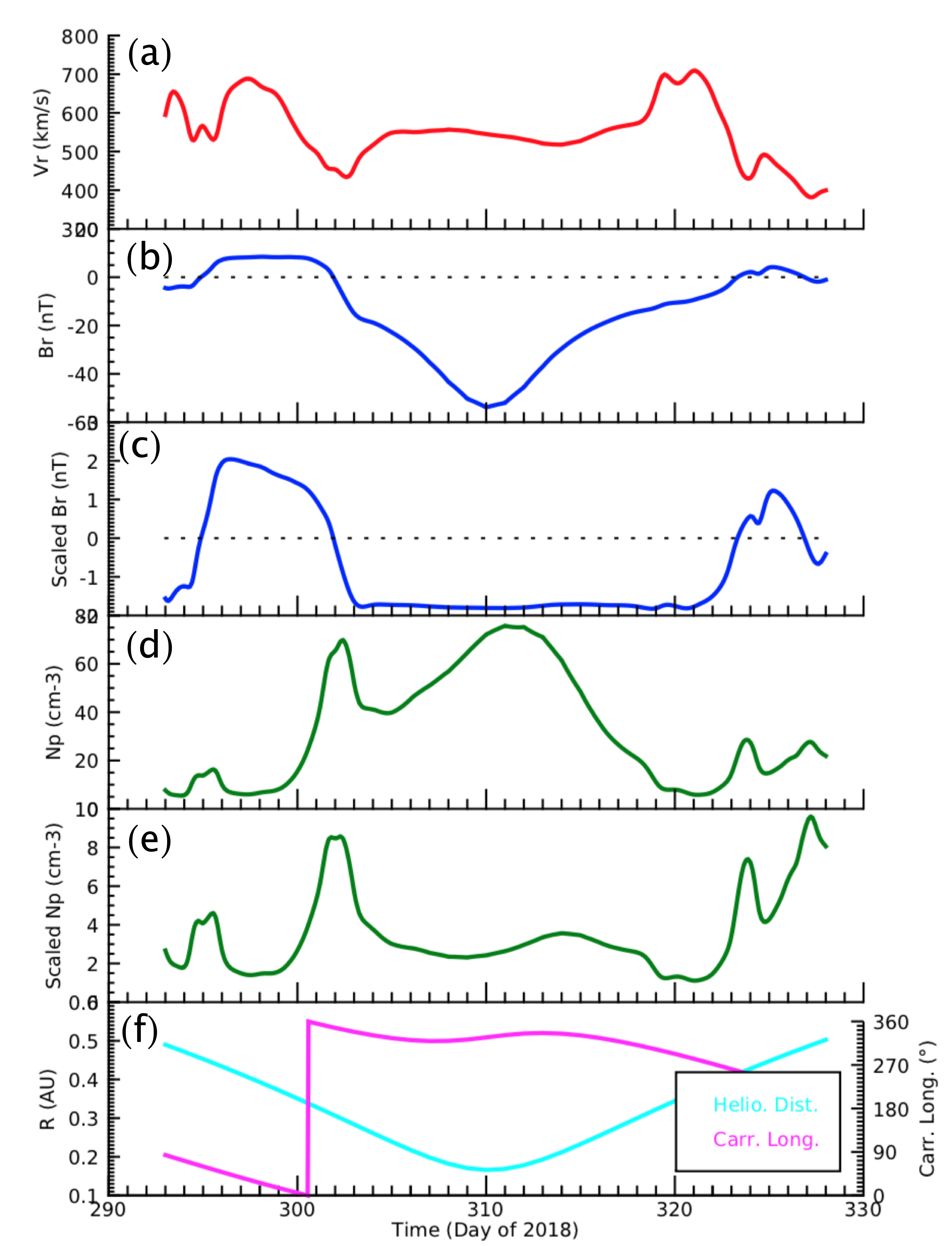}
\caption{Time series of predicted: (a) radial velocity ($v_r$); (b) radial magnetic field ($B_r$); (c) scaled $B_r$, (d) plasma number density ($N_p$); (e) scaled $N_p$, and (f) location of spacecraft in terms of heliocentric distance (aqua) and Carrington longitude (magenta).  In (c) and (e), the radial magnetic field and number density have been scaled by $r^2$ to account for the radial fall-off expected from flow into a spherical geometry.}
\label{fig-insitu}
\end{figure}

By flying PSP's trajectory through the heliospheric solution, we are able to recover the predicted plasma and magnetic field variables {\it in situ}. These are summarized in Figure~\ref{fig-insitu}. The panels show, from top to bottom: radial speed, radial component of the magnetic field ($B_r$), scaled $B_r$, plasma number density ($N_p$), scaled $N_p$, and the heliocentric position of the spacecraft, as summarized by distance ($R$) and Carrington longitude. Focusing first on panels (a), (b), and (d), we note the following points. First, the speed remained approximately constant during the interval from first encounter to second encounter ($\sim 525$ kms$^{-1}$. Second, the magnetic field switched from positive polarity to negative polarity (day of year: 302), and remained so for approximately three weeks. It also increased to a maximum negative value of -80 nT. Third, plasma density rose to almost 80 cm$^{-3}$; however, the peak in density lagged the magnetic field peak by approximately one day.  

These variations, however, are misleading, in the sense that both the density and radial magnetic field fluctuations lie on top of the geometric expansion of the solar wind, which naturally produces a $1/r^2$ fall off with heliocentric distance, or, since the spacecraft is moving towards perihelion, and increase of $r^2$. To address this, in panels (c) and (e) of Figure~\ref{fig-insitu}, we have removed the $1/r^2$ variation and scaled the measurements to 1 AU values. Both the values and variations are markedly different (compare panels (b) and (c) and panels (d) and (e)). Now, $B_r$ remains virtually flat ($\sim 8$ nT) for the entire duration of the encounter. Number density, $N_p$, on the other hand does have some intrinsic variations, peaking four days after perihelion (doy 314). This explains why the peaks in density and magnetic field are not synchronous. The peak in field is due to the spacecraft reaching perihelion, i.e., it is an expansion effect, whereas the density is a superposition of this expansion effect with a later, intrinsic peak due to structure generated by the Sun. 

\section{Conclusions and Discussion} \label{sec:conclusions}

Based on the analysis presented here, we conclude that in the days prior to the first encounter PSP was connected to field lines originating in the well-established positive-polarity, northern polar coronal hole. During the encounter, and at perihelion in particular, field lines at the spacecraft mapped back to a negative equatorial coronal hole. After the encounter (beyond doy 320), field lines were all connected to a positive-polarity, equatorial coronal hole to the east. 

Our comparison of the magnetic mappings based on the MHD and PFSS solutions revealed some important differences. Although these differences have been discussed in the past, where it was shown that the MHD solutions generally yielded more accurate results \citep{riley06b}, we cannot make that claim here; at least yet. Until the magnetic field measurements (and, to a lesser extent, the plasma moments) have been sent back and fully analyzed, it is quite possible that the PFSS model is correct. It will be interesting to see whether the observations are consistent with a crossing of the current sheet on doy 302 (10/29/18) as suggested by our MHD results, which was not predicted by the PFSS solutions. 

It is interesting to compare our predictions with those of \citet{holst19a}. They also predicted a crossing of the HCS; however, in their model, this occurred at two points, 11/03/2018 and 11/08/2019, with the interval between them (and containing perihelion) being of negative polarity. In contrast, our model only predicted one crossing, occurring on 10/29/2018. Additionally, at perihelion, they predicted lower plasma speeds (360 km s$^{-1}$ compared to our prediction of 525  km s$^{-1}$), higher densities (500 cm$^{-3}$ compared to 80 cm$^{-3}$) and smaller radial magnetic field strengths (28 nT compared to 80 nT). Once the measurements from PSP become available, understanding these differences could be a crucial step for improving the accuracy of these models. 

The model implemented here relied on a number of assumptions and approximations that might impact the accuracy of the results presented here. First, although the model used is time-dependent, in the sense of solving the time-dependent equations of MHD, the boundary does not evolve in time, thus, the solutions are ``quasi-stationary''. Second, these boundary conditions are constructed from Earth-based observations of the photospheric magnetic field. Since PSP was approximately 90 degrees to the west of Earth at the time of perihelion, the magnetic field lying underneath it at that time had been observed approximately seven days earlier (and shifted to the appropriate longitude in the Carrington map). Thus, the data used to derive the prediction at perihelion was approximately one week old. During this time photospheric fields can evolve significantly, as suggested by comparisons between CRs 2208, 2209, and 2210. 

The analysis presented here highlights the value that global MHD models can contribute to missions like PSP. In particular, they provide a global context for interpreting in situ measurements made by the FIELDS and SWEAP instruments. Moreover, they can be used support or refute theories of coronal heating and/or acceleration of the solar wind. Finally, they can be used to connect solar source regions to the structures and events measured in situ by the spacecraft. This will be particularly useful during observations of energetic particles \citep[e.g.][]{schwadron15a}. 

\acknowledgments

The authors gratefully acknowledge support from NASA (80NSSC18K0100, NNX16AG86G, 80NSSC18K1129, and 80NSSC18K0101), NOAA (NA18NWS4680081), and the U.S. Air Force (FA9550-15-C-0001).


\begin{thebibliography}{}
\expandafter\ifx\csname natexlab\endcsname\relax\def\natexlab#1{#1}\fi
\providecommand{\url}[1]{\href{#1}{#1}}

\bibitem[{{Aschwanden} {et~al.}(2008){Aschwanden}, {Burlaga}, {Kaiser}, {Ng},
  {Reames}, {Reiner}, {Gombosi}, {Lugaz}, {Manchester}, {Roussev}, {Zurbuchen},
  {Farrugia}, {Galvin}, {Lee}, {Linker}, {Miki{\'c}}, {Riley}, {Alexander},
  {Sandman}, {Cook}, {Howard}, {Odstr{\v c}il}, {Pizzo}, {K{\'o}ta}, {Liewer},
  {Luhmann}, {Inhester}, {Schwenn}, {Solanki}, {Vasyliunas}, {Wiegelmann},
  {Blush}, {Bochsler}, {Cairns}, {Robinson}, {Bothmer}, {Kecskemety},
  {Llebaria}, {Maksimovic}, {Scholer}, \&
  {Wimmer-Schweingruber}}]{aschwanden08a}
{Aschwanden}, M.~J., {Burlaga}, L.~F., {Kaiser}, M.~L., {et~al.} 2008, Space
  Sci. Rev., 136, 565

\bibitem[{Bale {et~al.}(2016)Bale, Goetz, Harvey, Turin, Bonnell,
  Dudok{\'a}de{\'a}Wit, Ergun, MacDowall, Pulupa, Andr{\'e},
  {et~al.}}]{bale16a}
Bale, S., Goetz, K., Harvey, P., {et~al.} 2016, Space science reviews, 204, 49

\bibitem[{Caplan {et~al.}(2017)Caplan, Miki{\'c}, Linker, \&
  Lionello}]{caplan17b}
Caplan, R.~M., Miki{\'c}, Z., Linker, J.~A., \& Lionello, R. 2017in , IOP
  Publishing, 012016

\bibitem[{Downs {et~al.}(2016)Downs, Lionello, Miki{\'c}, Linker, \&
  Velli}]{downs16a}
Downs, C., Lionello, R., Miki{\'c}, Z., Linker, J.~A., \& Velli, M. 2016, The
  Astrophysical Journal, 832, 180

\bibitem[{Fox {et~al.}(2016)Fox, Velli, Bale, Decker, Driesman, Howard, Kasper,
  Kinnison, Kusterer, Lario, {et~al.}}]{fox16a}
Fox, N., Velli, M., Bale, S., {et~al.} 2016, Space Science Reviews, 204, 7

\bibitem[{Kasper {et~al.}(2016)Kasper, Abiad, Austin, Balat-Pichelin, Bale,
  Belcher, Berg, Bergner, Berthomier, Bookbinder, {et~al.}}]{kasper16a}
Kasper, J.~C., Abiad, R., Austin, G., {et~al.} 2016, Space Science Reviews,
  204, 131

\bibitem[{Linker {et~al.}(1999)Linker, {Miki{\'c}}, Bisecker, Forsyth, Gibson,
  Lazarus, Lecinski, Riley, Szabo, \& Thompson}]{linker99a}
Linker, J.~A., {Miki{\'c}}, Z., Bisecker, D.~A., {et~al.} 1999, J. Geophys.
  Res., 104, 9809

\bibitem[{{Lionello} {et~al.}(2013){Lionello}, {Downs}, {Linker},
  {T{\"o}r{\"o}k}, {Riley}, \& {Miki{\'c}}}]{lionello13a}
{Lionello}, R., {Downs}, C., {Linker}, J.~A., {et~al.} 2013, Ap. J., 777, 76

\bibitem[{Lionello {et~al.}(2001)Lionello, Linker, \& Miki\'c}]{lionello01a}
Lionello, R., Linker, J.~A., \& Miki\'c, Z. 2001, Astrophys. J., 546, 542

\bibitem[{{Lionello} {et~al.}(2014){Lionello}, {Velli}, {Downs}, {Linker},
  {Miki{\'c}}, \& {Verdini}}]{lionello14a}
{Lionello}, R., {Velli}, M., {Downs}, C., {et~al.} 2014, Ap. J., 784, 120

\bibitem[{McComas {et~al.}(2016)McComas, Alexander, Angold, Bale, Beebe,
  Birdwell, Boyle, Burgum, Burnham, Christian, {et~al.}}]{mccomas16a}
McComas, D., Alexander, N., Angold, N., {et~al.} 2016, Space Science Reviews,
  204, 187

\bibitem[{Miki\'c \& Linker(1994)}]{mikic94a}
Miki\'c, Z., \& Linker, J.~A. 1994, Astrophys. J., 430, 898

\bibitem[{Mikic {et~al.}(2018)Mikic, Lionello, Downs, Linker, P., Shen, \&
  Raymond}]{mikic18a}
Mikic, Z., Lionello, R., Downs, C., {et~al.} 2018, in Solar Wind 15, ed.
  G.~Lapenta (Brussels, Belgium: AIP)

\bibitem[{{Parker}(1958)}]{parker58a}
{Parker}, E.~N. 1958, Astrophys. J., 128, 664

\bibitem[{Pizzo \& Gosling(1994)}]{pizzo94a}
Pizzo, V.~J., \& Gosling, J.~T. 1994, Geophys. Res. Lett., 21, 2063

\bibitem[{Riley {et~al.}(2001)Riley, Linker, \& Miki\'c}]{riley01a}
Riley, P., Linker, J.~A., \& Miki\'c, Z. 2001, J. Geophys. Res., 106, 15889

\bibitem[{{Riley} {et~al.}(2006){Riley}, {Linker}, {Miki{\'c}}, {Lionello},
  {Ledvina}, \& {Luhmann}}]{riley06b}
{Riley}, P., {Linker}, J.~A., {Miki{\'c}}, Z., {et~al.} 2006, Astrophys. J.,
  653, 1510

\bibitem[{{Riley} {et~al.}(2012){Riley}, {Lionello}, {Linker}, {Mikic},
  {Luhmann}, \& {Wijaya}}]{riley12a}
{Riley}, P., {Lionello}, R., {Linker}, J.~A., {et~al.} 2012, Solar Phys., 274,
  361

\bibitem[{{Schwadron} {et~al.}(2015){Schwadron}, {Lee}, {Gorby}, {Lugaz},
  {Spence}, {Desai}, {T{\"o}r{\"o}k}, {Downs}, {Linker}, {Lionello},
  {Miki{\'c}}, {Riley}, {Giacalone}, {Jokipii}, {Kota}, \&
  {Kozarev}}]{schwadron15a}
{Schwadron}, N.~A., {Lee}, M.~A., {Gorby}, M., {et~al.} 2015, Ap. J., 810, 97

\bibitem[{T{\"o}r{\"o}k {et~al.}(2018)T{\"o}r{\"o}k, Downs, Linker, Lionello,
  Titov, Miki{\'c}, Riley, Caplan, \& Wijaya}]{torok18a}
T{\"o}r{\"o}k, T., Downs, C., Linker, J.~A., {et~al.} 2018, The Astrophysical
  Journal, 856, 75

\bibitem[{{van der Holst} {et~al.}(2019){van der Holst}, {Manchester}, {Klein},
  \& {Kasper}}]{holst19a}
{van der Holst}, B., {Manchester}, IV, W.~B., {Klein}, K.~G., \& {Kasper},
  J.~C. 2019, Ap. J. Lett., 872, L18

\bibitem[{{Verdini} \& {Velli}(2007)}]{verdini07a}
{Verdini}, A., \& {Velli}, M. 2007, Astrophys. J., 662, 669

\bibitem[{Vourlidas {et~al.}(2016)Vourlidas, Howard, Plunkett, Korendyke,
  Thernisien, Wang, Rich, Carter, Chua, Socker, {et~al.}}]{vourlidas16a}
Vourlidas, A., Howard, R.~A., Plunkett, S.~P., {et~al.} 2016, Space Science
  Reviews, 204, 83

\bibitem[{{Zank} {et~al.}(1996){Zank}, {Matthaeus}, \& {Smith}}]{zank96a}
{Zank}, G.~P., {Matthaeus}, W.~H., \& {Smith}, C.~W. 1996, J. Geophys. Res.,
  101, 17093

\end{thebibliography}
\end{document}